\documentclass[twocolumn,english,prl,amsmath,amssymb,superscriptaddress,preprintnumbers]{revtex4-1}
\usepackage[utf8]{inputenc}
\usepackage{color}
\usepackage{amsmath}
\usepackage{amssymb}
\usepackage{graphicx}
\usepackage{textcomp}
\usepackage{dcolumn}
\usepackage{bm}
\usepackage[right]{eurosym}
\usepackage{float}
\usepackage[english]{babel}
\usepackage{blindtext}
\usepackage{babel}
\usepackage[normalem]{ulem}

\begin{document}

\title{Ignition of a helium nanoplasma by x-ray multiple ionization of a heavy rare-gas core}




\author{D. Schomas} 
\affiliation{Physikalisches Institut, Universit{\"a}t Freiburg, 79104 Freiburg, Germany}

\author{C. Medina} 
\affiliation{Physikalisches Institut, Universit{\"a}t Freiburg, 79104 Freiburg, Germany}

\author{L. Ben Ltaief} 
\affiliation{Department of Physics and Astronomy, Aarhus University, 8000 Aarhus C, Denmark}

\author{R. B. Fink} 
\affiliation{Department of Physics and Astronomy, Aarhus University, 8000 Aarhus C, Denmark}

\author{S. Mandal} 
\affiliation{Indian Institute of Science Education and Research, Pune 411008, India}

\author{S. R. Krishnan} 
\affiliation{Indian Institute of Technology Madras, Chennai 600036, India}

\author{R. Michiels}
\affiliation{Physikalisches Institut, Universit{\"a}t Freiburg, 79104 Freiburg, Germany}

\author{M. Debatin}
\affiliation{Physikalisches Institut, Universit{\"a}t Freiburg, 79104 Freiburg, Germany}
\altaffiliation[Present address: ] {Institut f{\"u}r Physik, Universit{\"a}t Kassel, 34132 Kassel, Germany} 

\author{F. Stienkemeier}
\affiliation{Physikalisches Institut, Universit{\"a}t Freiburg, 79104 Freiburg, Germany}

\author{S. Toleikis}
\affiliation{Deutsches Elektronen-Synchrotron, DESY, 22607 Hamburg, Germany}

\author{C. Passow}
\affiliation{Deutsches Elektronen-Synchrotron, DESY, 22607 Hamburg, Germany}

\author{N. Ekanayake}
\affiliation{Deutsches Elektronen-Synchrotron, DESY, 22607 Hamburg, Germany}

\author{C. Ott}
\affiliation{Max-Plank-Institut f{\"u}r Kernphysik, 69117 Heidelberg, Germany}

\author{R. Moshammer}
\affiliation{Max-Plank-Institut f{\"u}r Kernphysik, 69117 Heidelberg, Germany}

\author{T. Pfeifer}
\affiliation{Max-Plank-Institut f{\"u}r Kernphysik, 69117 Heidelberg, Germany}

\author{A. Heidenreich} 
\affiliation{Kimika Fakultatea, Euskal Herriko Unibertsitatea (UPV/EHU) and Donostia International Physics Center (DIPC), P.K. 1072, 20080 Donostia, Spain}
\affiliation{IKERBASQUE, Basque Foundation for Science, 48011 Bilbao, Spain}

\author{M. Mudrich} 
\affiliation{Department of Physics and Astronomy, Aarhus University, 8000 Aarhus C, Denmark}
\email[E-mail me at: ]{mudrich@phys.au.dk}

\begin{abstract}
The dynamics of an x-ray-ionized two-component core-shell nanosystem is probed using doped helium (He) nanodroplets.  First, a soft x-ray pump pulse selectively inner-shell ionizes the core cluster formed of heavier rare-gas atoms, causing electron migration from the He shell to the highly-charged core. This ignites a He nanoplasma which is then driven by an intense near-infrared probe pulse. The ultrafast charge redistribution, evidenced by the rise of He$^+$ and He$^{2+}$ ion yields from the nanoplasma within $\lesssim 70$~fs, leads to strong damping of the core cluster expansion. Thus, He droplets act as efficient tampers that reduce the radiation damage of embedded nanostructures, a property that could be exploited for improving coherent diffraction images.


\end{abstract}

\date{\today}

\maketitle

Free nanometer-sized atomic clusters and nanodroplets are attractive test objects for studying the intricate interaction of photons, electrons, and ions in strongly driven, highly correlated regimes of light-matter interaction. Detailed knowledge of the interaction processes occurring inside highly ionized nanosystems, so-called nanoplasmas, is essential for predicting and controlling the evolution of targets subjected to x-ray coherent diffraction imaging (CDI)~\cite{Gnodtke:2009,Gorkhover:2016,Fluckiger:2016}. Recording highly resolved coherent diffraction images of isolated nanoparticles and even individual molecules by single ultrashort x-ray pulses is one of the primary goals of existing and upcoming x-ray free electron lasers (FELs)~\cite{Neutze:2000,Chapman:2011,Barty:2013}. 

Helium nanodroplets are particularly well suited targets for studying the nanoplasma dynamics of heterogeneous systems due to their unique properties: He droplets efficiently pick up impurity atoms or molecules of any species; owing to the superfluid nature of the He droplets, these dopants are highly mobile inside a He nanodroplet where they aggregate into a cluster which is spatially and energetically well separated from the He environment. Since He droplets are transparent from the infrared up to extreme ultraviolet (XUV) spectral regions, dopants can be selectively excited or ionized without directly exciting the He host droplet~\cite{Toennies:2004,Stienkemeier:2006,Mudrich:2014}. 

When He nanodroplets are exposed to intense near-infrared (NIR) laser pulses, they ionize in an avalanche-like process, where the presence of dopants in the droplets drastically lowers the intensity threshold for ionization. The evolution and systematics of dopant-induced, strong-field NIR-driven He nanoplasmas have been studied in detail~\cite{KrishnanPRL:2011,Krishnan:2012,Heidenreich:2016,Heidenreich:2017,Kelbg:2018,Kelbg:2019}. In doped He nanodroplets, seed electrons were created by tunnel ionization of the dopants, and the He shell is ignited by laser-driven electron impact ionization (EII)~\cite{MikaberidzePRL:2009}. 

In the present work, we exploit for the first time another unique property of He nanodroplets – their near transparency to x-ray radiation. Using short soft x-ray pulses, we inner-shell photoionize heavier rare-gas atoms aggregated inside the nanodroplets without significantly photoionizing the surrounding shell of He atoms. In this way, a core of highly charged ions and a distribution of electrons is created, which ignite a He nanoplasma driven by an intense NIR pulse. The measured ultrafast build-up of the He nanoplasma ion yields in our soft x-ray-NIR pump-probe experiment reflects the ultrafast electron transfer from the He shell to the dopant core, as revealed by molecular dynamics (MD) simulations. While this electron transfer causes the He shell to become partly charged and to expand, the dopant core is rapidly neutralized and its expansion is damped. Thus, we demonstrate that He droplets serve as tampers that efficiently spatially confine x-ray ionized nanoclusters. This property of He nanodroplets could be exploited for improving the resolution of single-shot x-ray coherent diffraction images of embedded molecules and nanoparticles~\cite{Hoener:2008,Jurek:2008,Gnodtke:2009,Hau-Riege:2010,Ziaja:2011}. 

The experiments were performed at the XUV-FEL facility FLASH at DESY in Hamburg. A continuous He nanodroplet apparatus including a gas doping unit and a combined electron velocity-map imaging and ion time-of-flight mass spectrometer~\cite{Schomas:2017} was mounted at the open-port beamline BL2. The focused soft x-ray (5~nm) and NIR (800~nm) beams were collinearly superimposed using a mirror with a centered hole mounted at the exit of the beamline. The focal spot sizes ($1/e^2$ radius) of the x-ray and NIR beams in the interaction region of the spectrometer were $20$ and $50~\mu$m, respectively. The FWHM pulse durations were about $110$ and 55~fs and pulse energies were $75~\mu$J and up to $10$ mJ. In the experiments discussed here, the peak intensity of the x-ray pulses was $2.1\times 10^{13}$~Wcm$^{-2}$ and that of the NIR pulses was $1.8\times 10^{14}$~Wcm$^{-2}$. The delay time between the two pulses was controlled by a mechanical delay stage placed in the NIR laser beam. 

The mean sizes of the He nanodroplets were varied in the range $\langle N\rangle =4000$-$11000$ He atoms per droplet by varying the temperature of the He nozzle. The mean number of heavier rare-gas dopants embedded in the He droplets was controlled by the dopant partial pressure in the doping cell~\cite{Toennies:2004}. When He droplets are multiply doped by rare gas atoms, the dopants form clusters inside the He droplets, of which the binding energy is released into the droplets~\cite{Lewerenz:1995}. The droplets subsequently shrink by evaporation of He atoms. All droplet sizes specified in this work refer to the estimated reduced sizes due to dopant aggregation~\cite{Kuma:2007}.

He nanodroplets are nearly transparent at a wavelength of $5$~nm where the photoionization cross section of He (0.032~Mbarn) is by two orders of magnitude smaller than that of Ar (4.8~Mbarn), Kr (5.4~Mbarn) and Xe (1.6~Mbarn). Therefore, the yield of He ions from pure He nanodroplets irradiated by only the x-ray pulses fell below the detection threshold. Likewise, the NIR laser pulses alone did not substantially ionize pure He nanodroplets due to the high ionization energy of He. When doping the He droplets with small amounts of the heavier rare gases Ar, Kr, or Xe, the NIR pulse alone yielded only low rates of He$^+$ and He$_2^+$ ions. However, when applying both the x-ray and the NIR pulses simultaneously to doped He nanodroplets, large yields of He ions and electrons were detected.

Fig. 1 (a) shows a series of delay-dependent measurements of the yields of He$^+$ ions (symbols) for various numbers of Ar atoms embedded in He droplets of average size before doping $\langle N\rangle =8000$. The constant signal offset due to ionization by only the NIR pulse was subtracted. 
The delay-dependent yields of the most prominent ions, He$^+$, as well as the yields of He$^{2+}$, He$^{+}_2$ and of electrons follow the same characteristic pump-probe delay dependence. Likewise, the pump-probe transients recorded for krypton (Kr) and xenon dopants (not shown) display the same structure as those for Ar-doped He droplet. Fig. 1 (b) shows the He$^+$ yield measured for Kr-doped He droplets at various He droplet sizes. Followed by a steep rise within the temporal overlap of the two pulses, the He$^+$ yields reach a maximum value between 100 and 200~fs, and fall off within about 1~ps. Note that no delay dependence is measured for pure He droplets [Fig. 1 (a), Ar$_0$]. 

To extract the characteristic features from each delay-dependent measurement, the data are fitted by the model
\[ s(\Delta t)=s_0[\Theta(\Delta t)\exp\left(- \ln 2 \,\Delta t/\tau_2 \right)]\circ g(\Delta t,\tau_1), \]
where $g(\Delta t,\tau_1)=\exp\left(-\ln 2 \,\Delta t^2/\tau_1^2 \right)\sqrt{\ln 2/\pi}/\tau_1$ is a normalized Gaussian function. This model is rationalized by assuming that two infinitesimally short pulses induce a step-like signal rise $\Theta(\Delta t)$ at the delay $\Delta t=0$, followed by an exponential signal decay $\propto\exp\left(-\ln 2 \,\Delta t/\tau_2 \right)$. The finite pulse durations are taken into account by convolving this kernel with $g(\Delta t,\tau_1)$. Thus, in this model the rise time of the measured signal, given by $\tau_1$, should be determined by the width of the cross-correlation function of the two laser pulses, whereas $\tau_2$ determines the intrinsic decay time of the system. $s_0$ denotes the height of the signal maximum. 

\begin{figure}
	\center
	\includegraphics[width=1.05\columnwidth]{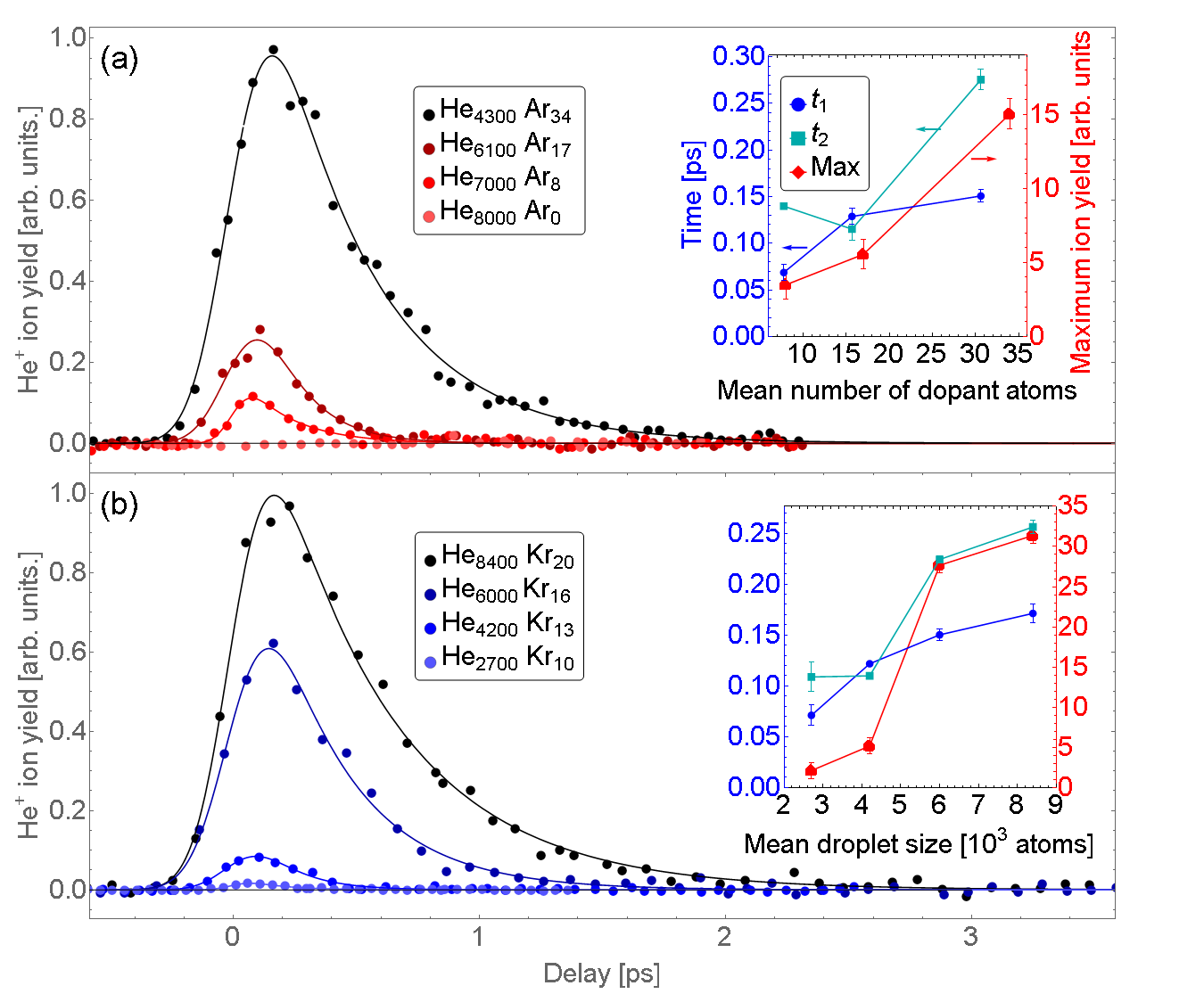}\caption{\label{fig1} He ion yields measured as a function of the delay between the x-ray pump and the NIR probe pulses. In panel (a), the He droplets initially consisting of about 8000 He atoms were doped with various numbers of Ar atoms. In (b), He droplets of variable size were doped with Kr atoms as indicated. The insets show the fit results for the rise times $\tau_1$, decay times $\tau_2$, and maxima of the He$^+$ yields.}
\end{figure}

Fig.~\ref{fig1} displays these fit results as insets in panels (a) and (b). The general trends can be summarized as follows: The ion yields significantly rise for increasing droplet size and doping level, as well as for rising FEL and NIR pulse intensities (not shown). 
The two characteristic times $\tau_1$ and $\tau_2$ roughly double in value over the varied ranges of droplet size and Ar doping level. Surprisingly, the shortest measured rise time of $\tau_1 = 70~$fs, measured for small droplets and low doping level, is significantly shorter than the width of the cross correlation of the two laser pulses, $\tau_\mathrm{corr}=130\pm 20$~fs. This finding leads us to conclude that the detected signal depends nonlinearly on the pulse intensities, where the signal is strongly enhanced in the delay range when the maxima of the two pulses nearly coincide. This is in line with the concept that nanoplasmas form in an avalanche-like ionization process~\cite{Fennel:2010,DoeppnerPRL:2010,Schuette:2016}. The ultrashort rise time of the ion yields indicates that small doped He droplets are maximally activated by secondary ionizations within the duration of the x-ray pulse, whereas larger He droplets are activated about 100~fs after the maximum of the x-ray pulse. A similar observation, recently made for pure Xe clusters irradiated by hard x-ray pulses~\cite{Kumagai:2018}, was interpreted by the interplay between electron-impact excitation and depletion of excited states by interatomic Coulombic decay (ICD). 

To obtain deeper insights, we perform systematic molecular dynamics (MD) simulations using the methodology described elsewhere~\cite{Heidenreich:2016,Heidenreich:2017}.
The x-ray ionization is implemented in an \textit{ad hoc} procedure, where all Ar atoms are singly 3p or 2p photoionized, the latter triggering the emission of one or two Auger electrons with branching ratios reproducing the experimental relative charge state abundances. These abundances ($0.04$ Ar$^+$, $0.80$ Ar$^{2+}$ and $0.16$ Ar$^{3+}$) are taken from our measurements of pure Ar atoms and agree well with previous results~\cite{Lablanquie:2007}. Photo and Auger ionizations are assumed to proceed instantaneously, generating a distribution of low and high-energy electrons.

\begin{figure}
	\center
	\includegraphics[width=0.9\columnwidth]{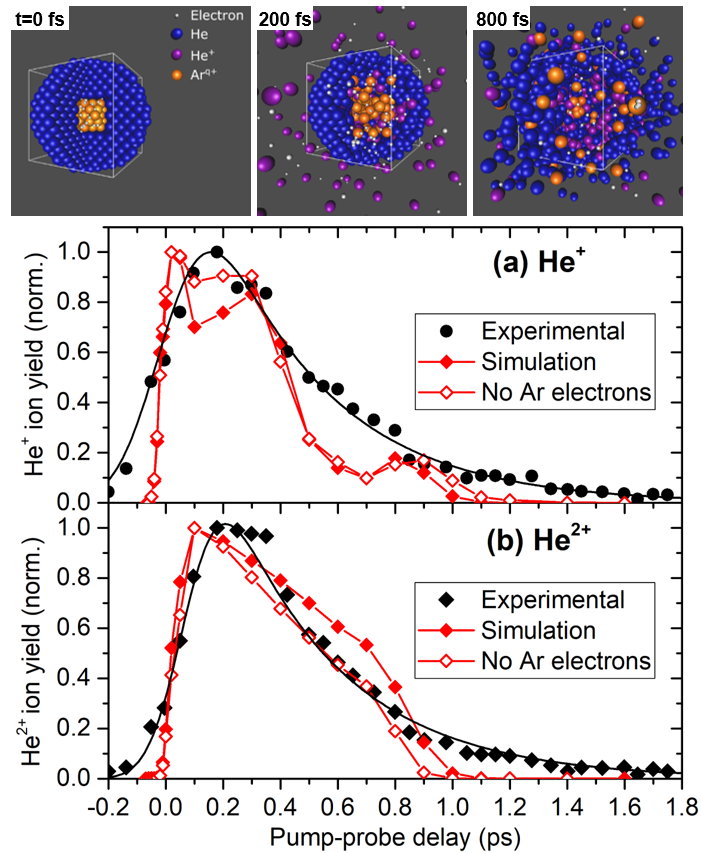}\caption{\label{fig2} Top row: Snapshots of the evolution of a He$_{2171}$Ar$_{55}$ cluster x-ray ionized at $t=0$. Balls of different color indicate free electrons, neutral and ionized He atoms, and Ar ions. Experimental and simulated He$^+$ (a) and He$^{2+}$ (b) ion yields as a function of the pump-probe delay.}
\end{figure}

Snapshots of a He$_{2171}$ droplet doped with an Ar$_{55}$ cluster are shown on the top of Fig.~\ref{fig2}. The Ar nanoplasma created at $t=0$ (left panel) induces 100 more Ar and 300 He ionizations during the first 200-300~fs, before the arrival of the NIR pulse. Most of these He$^+$ ions exit the droplet within 200~fs (middle panel). Neutral He atoms, which acquire kinetic energy by electron-atom collisions, as well as Ar ions expand more slowly (right panel). When the NIR probe pulse arrives at $\Delta t\gtrsim 1$~ps, the droplet is so thinned out that ionization by the NIR pulse is suppressed (not shown). In contrast, at shorter delay the droplet is still sufficiently compact that the He shell avalanche ionizes and subsequently Coulomb explodes.

A comparison of experimental and simulated x-ray-NIR pump-probe ion yields He$^+$ and He$^{2+}$ is displayed in Fig.~\ref{fig2} (a) and (b), respectively, for He droplets of size $\langle N\rangle =4300$ doped with 34 Ar atoms. The simulated pump-probe curves are obtained for the He$_{2171}$Ar$_{55}$ system. For each delay, the signal is averaged over 300 MD trajectories and over the NIR laser intensities in the focal volume similarly to Ref.~\cite{Heidenreich:2011}. 

To test the response of the He shell in the extreme case where all photo and Auger electrons instantaneously leave the droplet at $t=0$, simulations were carried out for the corresponding highly cationic Ar clusters (open symbols in Fig.~\ref{fig2}). 
Strikingly, the two simulations nearly coincide, indicating that the creation of multiply charged Ar ions inside the He nanodroplets is an efficient means of igniting a nanoplasma. Furthermore, it shows that a droplet of $2000$ He atoms is large enough to substantially stabilize even a multiply charged pure cation cluster. Only the He$^{2+}$ yield decays slightly faster when emitted electrons are removed, indicating that bare Ar$^{q+}$ ions are more efficient in ionizing the He shell which then expands faster. 

The rising edges in the simulations are steeper than in the experiment due to the assumption of an instantaneous x-ray ionization of the Ar core in the simulation. The drop of the simulated He$^+$ yield at $\Delta t>100$~fs is slightly faster than the measured one; likely, in the experiment, more larger He droplets out of the broad He droplet size distribution contribute which feature longer decay times $\tau_2$. Larger He droplets contain more dopants and are more susceptible to NIR strong-field ionization. Their  expansion induced by x-ray ionization of the core is delayed compared to small droplets. The ratio of He$^{2+}$ vs. He$^+$ yields (not shown) is slightly overestimated in the simulation ($0.40$) compared to the experiment ($0.25$). The overall good agreement of the experimental data with the results of the MD simulation shows that the nanoplasma evolution follows classical dynamics and  ICD processes need not be invoked, as opposed to previous findings~\cite{Kumagai:2018}.

\begin{figure}
	\center
	\includegraphics[width=1.0\columnwidth]{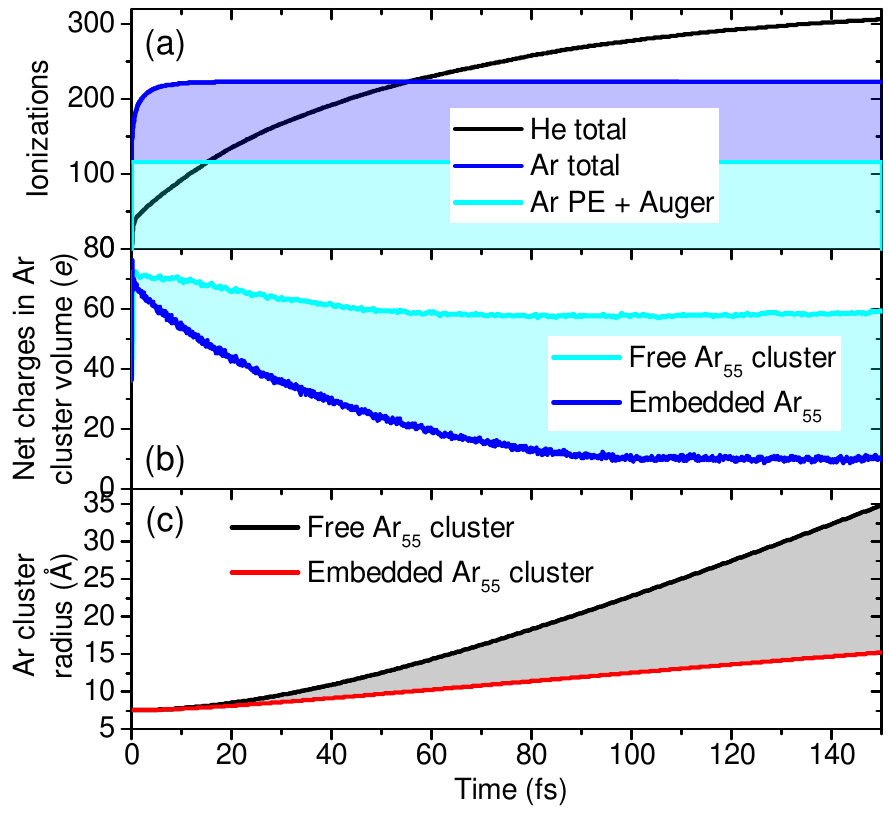}\caption{\label{fig3} (a) Time evolution of the number of Ar and He ionizations induced in a He$_{2171}$Ar$_{55}$ core-shell cluster by an x-ray pulse. (b) Sum of all charges inside the expanding Ar cluster volume for a free Ar$_{55}$ cluster (cyan line) and an He$_{2171}$Ar$_{55}$ cluster (blue line). (c) Expansion dynamics of a free Ar$_{55}$ cluster (black line) compared to an embedded Ar$_{55}$ cluster (red line).}
\end{figure}

In the remainder of this Letter, we focus on the short-time dynamics which determines the rising edge in the pump-probe traces and which is relevant for CDI. Fig.~\ref{fig3} shows several time-dependent quantities during the first 150 fs after x-ray ionization of an Ar$_{55}$ cluster embedded in a He$_{2171}$ droplet in the absence of the NIR pulse. Panel (a) depicts the number of inner ionizations~\cite{Last:2000} of the Ar and the He atoms in response to creating Ar$^{q+}$ and electrons. Initiated by the ionization of the Ar$_{55}$ cluster by the x-ray pulse at $t=0$ (cyan curve and area), the Ar is further ionized mostly by EII (blue line and area) within the first two fs, creating mostly Ar$^{4+}$ and Ar$^{5+}$ ions. Electrons diffusing from the Ar nanoplasma into the He shell generate He$^+$ and additional quasifree electrons by EII assisted by the electric field of the Ar ions (black solid line). This proceeds over a period of more than 200 fs owing to the large number of He atoms surrounding the Ar core. The large number of quasifree electrons created by the highly charged Ar$^{q+}$ can subsequently be driven by the NIR pulse to transform the whole core-shell cluster into a nanoplasma.

But how do the charges redistribute inside this core-shell system during the activation phase? This can be seen from the time-evolution of the total net charge $Q$ of the cluster core, \textit{i.~e.} the sum of all ions and electron charges inside the expanding Ar cluster volume. Fig.~\ref{fig3} (b) shows the evolution of $Q$ for an embedded Ar$_{55}$ cluster and, for comparison, for a free Ar$_{55}$ cluster (cyan line). The cyan shaded area shows that for the embedded Ar cluster, the initial Ar$^{q+}$ charges are substantially neutralized by electron transfer from the He shell to the Ar$_{55}$ cluster. 
This massive charge redistribution strongly affects the expansion dynamics of the system, as previously observed for heavier rare gas clusters irradiated by XUV pulses~\cite{Hoener:2008}. Fig.~\ref{fig3} (c) shows the radius of the Ar cluster, taken as the largest distance of an Ar atom from the Ar$_{55}$ center of mass. The embedded Ar cluster expands much more slowly (red line) compared to the free Ar cluster (black line), showing that the He droplet acts as a tamper that strongly dampens the expansion of the ionized core cluster. 
\begin{figure}
	\center
	\includegraphics[width=1.0\columnwidth]{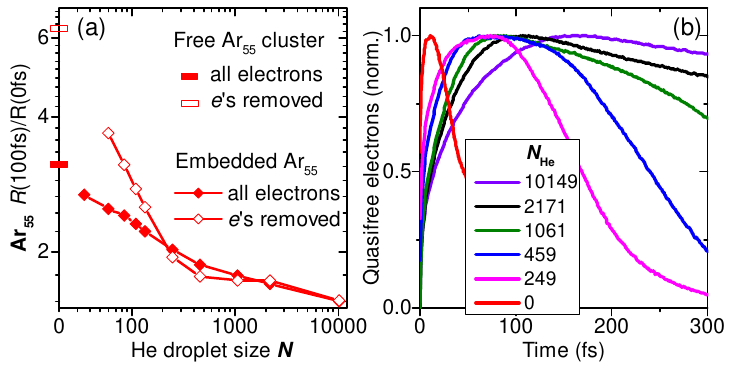}\caption{\label{fig4} (a) Dependence of the expansion of the x-ray-ionized Ar$_{55}$ cluster on the size of the He shell. Open symbols show the results of a simulation where the x-ray-emitted electrons are artificially removed. (b) He droplet size dependence of the evolution of the number of nanoplasma electrons in the initial volume of the droplet.}
\end{figure}
The tamper effect features a strong dependence on the size of the He droplet, as previously shown~\cite{Hoener:2008,Gnodtke:2009}. The increase of the Ar cluster radius at $t=100$~fs relative to its initial size is displayed in Fig.~\ref{fig4} (a). This factor amounts to 3.0 for a free Ar$_{55}$ cluster and to 6.2 for a free Ar cluster if the x-ray-emitted electrons are instantaneously removed, see the marks on the vertical scale. For a He droplet of size $N=10^4$ the expansion of the embedded Ar cluster within 100~fs is largely suppressed (ratio of radii $=1.6$).

The rise times $\tau_1$ of the pump-probe traces [insets in Fig.~\ref{fig1} (a) and (b)] can be rationalized by the charging dynamics of the x-ray-ionized Ar-He core-shell system. The NIR pulse optimally drives an ionization avalanche when a maximum of nanoplasma electrons has accumulated inside the nanodroplet. As the He shell is ionized by the diffusive motion of electrons in the field of the Ar ions in the core [Fig.~\ref{fig3} (a)], we expect that the maximum of the He ionization is reached at later times for larger He shells. Indeed, the number $N_e$ of nanoplasma electrons in the initial volume of the nanodroplet, shown in Fig.~\ref{fig4} (b), reaches a maximum that shifts in time from $t=70$ to $160$~fs when the He droplet size increases from 249 to 10149 He atoms per droplet. For better comparability, each $N_e(t)$ curve is normalized to its maximum. 
The drop of $N_e(t)$ at later times is due to the stagnation of EII by an interplay of decreasing electron kinetic energies and particle loss from the droplet as it starts to expand. The shifting of the $N_e(t)$ maxima is in good agreement with the shifting of the maxima of the measured pump-probe traces to longer delays, as quantified by $\tau_1$ (Fig.~\ref{fig1}). To directly probe the electron-transfer dynamics inside the nanodroplet and the resulting tampering of the core-cluster expansion, CDI experiments should be carried out using ultrashort x-ray pulses as they are becoming available from modern FELs such as the European XFEL.


The authors gratefully acknowledge financial support from the Basque Government (project IT1254-19) and from the Spanish Ministerio de Economia y Competividad (project CTQ2015-67660-P), Deutsche Forschungsgemeinschaft (DFG) within the project MU 2347/12-1 and STI 125/22-2 in the frame of the Priority Programme 1840 ‘Quantum Dynamics in Tailored Intense Fields’, and the Carlsberg Foundation. Computational and manpower support provided by IZO-SGI SG Iker of UPV/EHU and European funding (EDRF and ESF) is gratefully acknowledged. We would like to thank B. Manschwetus, R. Treusch, B. Erk and the whole FLASH team for assistance and support during the experiment, and the group of K.-H. Meiwes-Broer for loaning parts of the apparatus.


%


\end{document}